\documentclass[preprint,12pt]{elsarticle}

\usepackage{amssymb,amsmath}
\journal{Computational Materials Science}

\begin{document}

\begin{frontmatter}



\title{A Probabilistic Model for LCF}


\author{S. Schmitz${}^a$, T. Seibel${}^b$, T. Beck${}^c$, G. Rollmann${}^d$, R. Krause${}^e$, H. Gottschalk${}^f$}

\address{a Corresponding author: Universita
della Svizerra Italiana, Istituto di Scienze Computazionali, Via
Giuseppe Buffi 13, 6900 Lugano, Switzerland, email:
sebastian.schmitz@usi.ch, phone: +49 16095331082}
\address{b Institute of Energy and Climate Research (IEK-2), FZ J\"ulich, 52425 J\"ulich, Germany, email: t.seibel@fz-juelich.de}
\address{c Institute of Energy and Climate Research (IEK-2), FZ J\"ulich, 52425 J\"ulich, Germany, email: t.beck@fz-juelich.de}
\address{d Siemens AG Energy, Mellinghoferstra\ss e 55, 45473 M\"ulheim an
der Ruhr, Germany, email: georg.rollmann@siemens.com}
\address{e Universita
della Svizerra Italiana, Istituto di Scienze Computazionali, Via
Giuseppe Buffi 13, 6900 Lugano, Switzerland, email:
rolf.krause@usi.ch}
\address{f Bergische Universit\"at Wuppertal, Fachbereich
Mathematik und Naturwissenschaften, Gau\ss stra\ss e 20, 42097
Wuppertal, Germany, email: hanno.gottschalk@uni-wuppertal.de}

\begin{abstract}
Fatigue life of components or test specimens often exhibit a
significant scatter. Furthermore, size effects have a
non-negligible influence on fatigue life of parts with different
geometries.
%Under the same loading and temperature
%conditions a larger component is expected to have a shorter
%fatigue life.
We present a new probabilistic model for low-cycle fatigue (LCF)
in the context of polycrystalline metal. The model takes size
effects and inhomogeneous strain fields into account by means of
the Poisson point process (PPP). This approach is based on the
assumption of independently occurring LCF cracks and the
Coffin-Manson-Basquin (CMB) equation. Within the probabilistic
model, we give a new and more physical interpretation of the CMB
parameters which are in the original approach no material
parameters in a strict sense, as they depend on the specimen
geometry. Calibration and validation of the proposed model is
performed using results of strain controlled LCF tests of
specimens with different surface areas. The test specimens are
made of the nickel base superalloy RENE 80.
\end{abstract}

\begin{keyword}
Fatigue; Poisson point process; Coffin-Manson-Basquin equation

\end{keyword}

\end{frontmatter}

% \linenumbers


\section{Introduction}
\label{section_introduction}

In fatigue analysis, standardized specimen tests are commonly used
to represent temperature and stress conditions in engineering
parts under cyclic loading. Results of these tests are usually
visualized in E - N and S - N diagrams (W\"ohler curves) for
strain and stress controlled fatigue, respectively. Within the
safe-life approach of fatigue design, these diagrams are employed
to predict the fatigue life of engineering parts. In addition,
safety factors are applied to take the scatter for fatigue life,
size effects and uncertainties in the stress and temperature
fields into account. In contrast to that, in a probabilistic
approach, these quantities are explicitly taken into account: In
the present paper, such a probabilistic model is presented for the
failure mechanism of surface driven low-cycle fatigue (LCF).
Inhomogeneous strain fields as well as the size effect are
inherently considered within the model. As a consequence, this
opens up the possibility for a new, more physical approach to the
Coffin-Manson-Basqin (CMB) equation whose parameters in the
original interpretation are no material parameters in a strict
sense since they depend on the specimen geometry.

The stochastic nature of LCF crack initiation is a result of the
LCF failure mechanism on micro- and mesoscales, confer
\cite{Harders_Roesler,Fedelich,Sornette}. The residual scatter of
the number of load cycles to crack initiation is typically rather
large\footnote{Even under lab conditions, the factor between the
highest and lowest load cycles to crack initiation can easily
become 10 and higher, when 10-100 specimens are tested (a typical
number for industrial applications).} so that reliability
statistics is supposed to play an important part in LCF design. As
LCF cracks mostly initiate at the surface we focus on surface
driven LCF. For different materials and temperature regimes, there
are numerous physical mechanisms that can lead to the formation of
technical LCF cracks. However, during their initiation phase, LCF
cracks only influence strain fields on the micro- and mesoscales,
because they are very small in their spatial extent compared to
specimen or part dimensions. Therefore, one can assume that LCF
crack formation in one surface region has no impact on the crack
forming process on another part of the surface. In particular,
this is supported by the fact that we only consider the number of
cycles until the first crack has initiated. We can thus consider
crack formation as a problem of spatial statistics \cite{Sherman}
and use the notion of the Poisson point processes (PPP), confer
\cite{Kallenberg,Baddeley}. The corresponding intensity measure is
a surface integral over some local function that is supposed to
depend on the local strain field. This local function can also be
interpreted as a hazard density function \cite{Escobar_Meeker},
which constitutes another possible starting point for deriving
this probabilistic model. Note that \cite{Fedelich} also
emphasizes the role of the PPP.

For the density of the intensity measure we employ a Weibull
approach which is commonly used in reliability statistics, confer
\cite{Escobar_Meeker}. This results in a Weibull distribution for
the number $N$ of cycles to first crack initiation with a scale
parameter given by the CMB equation. In this context, our model
leads to a new interpretation of the CMB parameters which are now
independent of the geometry due to the incorporated size effect
via the assumption of independently occurring first crack
initiations. We also show a functional relationship between the
classical parameters for a specific geometry and those newly
interpreted and more physical ones. The model is calibrated by
means of LCF test results of standardized specimens and of the
maximum likelihood method \cite{Kallenberg}. After the
calibration, we predict W\"ohler curves for different specimen
geometries to validate our probabilistic model for LCF. Note that
it is in principle possible to calibrate our model with LCF test
results of arbitrary geometries\footnote{In order to apply
continuum mechanics the geometry has to consist of sufficiently
many grains. Moreover, information on when the first LCF crack
initiation occurred is needed which can be practically difficult.}
under surface driven LCF failure mechanism. Thus, notched
specimens or
even engineering parts %\footnote{Here, field data for the number of
%load cycles without any LCF crack initiations in the part is needed.}
in conjunction with finite element analysis (FEA) simulations
could be used for the calibration.

In the first part of Section \ref{section_theory} we recall the
CMB life prediction approach which will be the deterministic basis
of our probabilistic model. In Subsection
\ref{subsection_stochastics} we derive a model for crack
initiation based on the PPP which leads to the probabilistic model
for LCF presented in Subsection
\ref{subsection_stochastics_model}. At the end of this section, we
give a new interpretation of the CMB parameters. In Section
\ref{section_experiments} we discuss results of the experimental
validation of our model by considering different specimen
geometries. Section \ref{section_conclusion} ends this paper by
summarizing the theoretical and experimental results of this work
and by giving an outlook on future work on this probabilistic
model.

%%%%%%%%%%%%%%%%%%%%%%%%%%%%%%%%%%%%%%%%%%%%%%%%%%%%%%%%%%%%%%%%%%%%%%
\section{Fatigue and reliability statistics}\label{section_theory}

In this section, we discuss theoretical concepts in fatigue
analysis and then introduce our probabilistic model for LCF which
is based on the PPP. The model can also be
derived from a spatial hazard approach.%, confer \cite{Gottschalk_Schmitz}.

%%%%%%%%%%%%%%%%%%%%%%%%%%%%%%%%%%%%%%%%%%%%%%%%%%%%%%%%%%%%%%%%%%%%%%
\subsection{Fatigue and the Coffin-Manson-Basquin
Equation}\label{subsection_fatigue}

In fatigue analysis, the CMB equation is often used to describe
the relationship between strain or stress and the number of cycles
until crack initiation with respect to standardized test
specimens\footnote{Standardized test specimens are characterized
by a smooth and cylindrical shape, see Figure
\ref{Figure_specimens}.}, confer \cite{Harders_Roesler}. Here, we
focus on surface driven and strain controlled fatigue and refer to
\cite{Harders_Roesler,Fedelich,Sornette,Vormwald} for backgrounds
on fatigue failure mechanism with respect to polycrystalline
metal.
%\begin{figure}[t]
%    \centering
%\scalebox{0.28}{\input{Doktorarbeit_EN_diagram_2.pstex_t}}
%\caption{EN-DIAGRAM OF A STANDARDIZED SPECIMEN.}
%    \label{Figure_woehler_diagram_strain}
%\end{figure}

Following \cite{Harders_Roesler} the Basquin equation
$\varepsilon_a^{el}=\frac{\sigma_f'}{E}(2{N_i})^b$ can be used to
describe the range of a W\"ohler curve which is dominated by
elastic behavior. The parameter $\sigma_f'$ is called fatigue
strength coefficient and $b$ fatigue strength exponent. For the
plastic range of the W\"ohler curve the Coffin-Manson equation
$\varepsilon_a^{pl}=\varepsilon_f'(2{N_i})^c$ can be employed,
where the parameters $\varepsilon_f'$ and $c$ are called fatigue
ductility coefficient and fatigue ductility exponent,
respectively. Confer \cite{Sornette} for a detailed discussion of
the physical origin of this equation. Combining the previous
equations results in the CMB equation
\begin{equation}\label{materials_1.1}
\varepsilon_a=\frac{\sigma_f'}{E}(2{N_i})^b+\varepsilon_f'(2{N_i})^c,%=\varepsilon_a^{el}+\varepsilon_a^{pl}
\end{equation}
whose parameters can be estimated according to test data, confer
\cite{Escobar_Meeker,Schott} and Subsection
\ref{subsection_stochastics_model}.

In case of surface driven LCF, structural design concepts often
consider the component's position of highest surface strain and
then take the W\"ohler curve into account which corresponds to the
conditions at that surface position. In addition, safety factors
are introduced to account for the stochastic nature of fatigue,
for size effects\footnote{Note that different geometries of test
specimens lead to different W\"ohler curves, for example. We will
discuss size effects in more detail in Subsection
\ref{subsection_interpretation_CMB} and
\ref{subsection_calibration}.} and for uncertainties in the stress
and temperature fields. Note that sometimes more than one position
of highest surface strains is considered. This concept and
variants thereof are called safe-life approach to fatigue design
and they are often used in engineering, confer
\cite{Harders_Roesler}. In contrast to this approach, we now
present a statistical model for crack initiation.

%%%%%%%%%%%%%%%%%%%%%%%%%%%%%%%%%%%%%%%%%%%%%%%%%%%%%%%%%%%%%%%%%%%%%%
\subsection{Crack Initiation as Spatio-Temporal Poisson Point Process}\label{subsection_stochastics}

Our probabilistic model for LCF is based on a
%We now want to state the assumptions for our
statistical model for crack initiation which we present in this
subsection. To this aim, we introduce the variable $n$ of load
cycles that the component underwent. Though strictly speaking $n$
is a natural number, we follow the widely spread convention to
treat $n$ as a continuous, time-like number.

Let $\Omega$ be the volume filled by the mechanical component and
$\partial\Omega$ its surface. We assume that we can associate a
surface location $x$ in $\partial\Omega$ and a cycle number $n$
between zero and infinity to the initiation of each LCF crack. We
further consider collections $B$ of such pairs of locations and
times, i.e.\ $B\subset \partial\Omega\times (0,\infty]$ in
mathematical terms\footnote{Here and in the following some
mathematical details (e.g.\ measurability of and
$\sigma$-additivity in $B$) are deliberately suppressed.}. By
$N(B,\epsilon)$ we denote the number of cracks that initiated at
some location and time in $B$, given the strain state
$\varepsilon=\varepsilon(x)$ on the component's surface. As the
number of cracks initiating in a time interval in a certain
surface region is not predictable,
$N(B,\varepsilon)=N(B,\varepsilon)(\omega)$ is a random quantity.
Obviously, if $B$ can be decomposed into subsets $B_1,\ldots,B_n$
without any overlap, we have
\begin{equation}
\label{eqaAdditivity}
N(B,\varepsilon)=\sum_{j=1}^nN(B_j,\varepsilon).
\end{equation}
Hence the random counts of crack initiations with specified
location and time set $B$ have the structure of random counting
measures -- also called point processes -- that are studied
extensively in the mathematical literature \cite{Kallenberg}.

So far we have only set a mathematical frame. The assumptions
underlying our model are as follows
\begin{itemize}
\item[A1)] {\bf Identification of single cracks:} At one location and point in time,
at most one crack can initiate.
\item[A2)] {\bf Local dependence of the load situation:} Two surface regions with the same surface area and the same
strain state will have the same statistical properties of crack
initiation in any given time interval.
\item[A3)] {\bf Independence:} Given a number $B_1,\ldots,B_n$ of non-intersection collections of location and
time instances, the random counts $N(B_1,\varepsilon)$, $\dots$,
$N(B_n,\varepsilon)$ of cracks initiated in $B_1,\ldots,B_n$ are
statistically independent.
\end{itemize}
Let us discuss the above assumptions. The first one is just a
convention on what we consider to be a crack. The second
assumption just states that initiation of cracks is a local
phenomenon. In the given formulation, it also rules out some
potentially interesting effects - like dependency of the LCF crack
count statistics on local curvature of the surface. Also we
assumed constant material properties over the surface of the
component and constant temperature. It is however not difficult to
consider extended models with local temperature- and curvature
fields in addition to the strain field.

The last assumption is justified by the fact that, in the regime
that we consider, LCF cracks are sufficiently small such that they
do not change the strain state on a macroscopic state. Thus, the
initiation of a crack at some surface location and time does not
influence the initiation of crack at another time and another
location. This assumption will however break down at a subgranual
scale, as a LCF crack will traverse the entire grain. We however
consider this as a good approximation as long as the number of
grains on a surface with specified load is sufficiently large,
which in particular is the case in LCF material testing, confer
Section \ref{section_experiments}.

It is an interesting fact that the above three assumptions already
imply that $N(B,\varepsilon)$ is Poisson distributed
$N(B,\varepsilon)\sim{\rm Po}(\lambda(B,\varepsilon))$,
confer\footnote{Here assumption A1) encodes the mathematical
property of simplicity and A3) the independence of increments in
the language of that reference.} \cite[Corollary 7.4]{Kallenberg}.
Thus we have the following probabilities for the number of crack
initiations in $B$
\begin{equation}
\label{eqaPoisson}
P(N(B,\varepsilon)=r)=e^{-\lambda(B,\varepsilon)}\frac{\lambda(B,\varepsilon)^r}{r!},~~r=0,1,2\ldots~
.
\end{equation}
Here $\lambda=\lambda(B,\varepsilon)$ is the intensity parameter
of the Poisson distribution which is equal to the expected value,
denoted by $\mathbb{E}$, of crack initiation counts in $B$, given
the strain state $\varepsilon$ on the surface $\partial\Omega$ of
the component. Note that by (\ref{eqaAdditivity}) we have for the
expected values of crack initiation counts in $B$, given the
decomposition of $B$ into $B_1,\ldots,B_n$ as described above,
\begin{equation}
\lambda(B,\varepsilon)=\mathbb{E}\left[\sum_{j=1}^nN(B_j,\varepsilon)\right]=\sum_{j=1}^n\mathbb{E}\left[N(B_j,\varepsilon)\right]
=\sum_{j=1}^n\lambda(B_j,\varepsilon).
\end{equation}
Consequently, $\lambda(.,\varepsilon)$ is additive in the set
argument $B$. A reasonable model for $\lambda(.,\varepsilon)$
realizing assumptions A1) and A2) is given by
\begin{equation}
\label{eqaIntensity}
\lambda(B,\varepsilon)=\int_B\rho(n,\varepsilon) \,dAdn,
\end{equation}
with $dA$ the surface volume measure on $\partial\Omega$ and
$\rho(n,\varepsilon)$ the crack formation intensity function. The
latter carries the dimension av.\ counts of crack initiation per
square meter and load cycle.

If $B$ consists of all location and time instances with locations
in a portion $D$ of the surface $\partial \Omega$ and 'time' $n$
in some interval $(s,t]$, the Poisson intensity takes the form
$\lambda(D\times(s,t])=\int_s^t\int_D\rho(n,\varepsilon) \,dAdn$.

In particular, we are interested in the situation, where the
entire component is crack free up to some time (cycle number) $n$,
which we define as survival up to this time. Let $N_i$ be the
(random) time of initiation of the first crack on
$\partial\Omega$, i.e.\ $N_i$ equals the minimal $n$ such that
$N(\partial\Omega\times (0,n])>0$. Survival up to time $n$ is
defined as the absence of a crack up to that time. The probability
of survival up to time $n$, $S_{N_i}(n)=P(N_i>n)$, then is given
by
\begin{align}
\begin{split}
\label{eqaSurvival}
S_{N_i}(n)&=P(\mbox{'no crack initiation on $\partial\Omega$ up to $n$'}) \\
&=P(N(\partial\Omega\times(0,n],\varepsilon)=0)=\exp\left\{-\int_0^n\int_{\partial\Omega}\rho(\varepsilon,n)\,dAdn\right\}.
\end{split}
\end{align}

From (\ref{eqaSurvival}) we immediately deduce the following
expressions for the distribution function
\begin{equation}
\label{eqaDist}
F_{N_i}(n)=1-S_{N_i}(n)=1-\exp\left\{-\int_0^n\int_{\partial\Omega}\rho(\varepsilon,n)\,dAdn\right\}
\end{equation}
and hazard rate function
\begin{align}
\begin{split}
\label{eqaHazard}
h_{N_i}(n)&=\lim_{\Delta n\searrow 0}\frac{1}{\Delta n}P(N_i\in(n,n+\Delta n]|N_i>n)\\
&=\lim_{\Delta n\searrow 0}\frac{1}{\Delta n}\frac{F_{N_i}(n+\Delta n)-F_{N_i}(n)}{S_{N_i}(n)}\\
&=\int_{\partial\Omega}\rho(\varepsilon,n)\,dA.
\end{split}
\end{align}

%%%%%%%%%%%%%%%%%%%%%%%%%%%%%%%%%%%%%%%%%%%%%%%%%%%%%%%%%%%%%%%%%%%%%%%%%%%%%%%%%%%%

\subsection{A Probabilistic Model for LCF}\label{subsection_stochastics_model}

In this subsection, we introduce a model for the up to now
unspecified crack formation intensity function
$\rho(n,\varepsilon)$. Here, $\varepsilon$ is the strain field in
the geometry under consideration. In the case of a uniaxially
loaded specimen, this could be a constant, while for a more
complex part this can be obtained as the result of an FEA, confer
\cite{Finite_Elements_Ern,Finite_Elements_Ciarlet_1}. In this
work, we focus on polycrystalline metal such as RENE 80 which is
considered to be isotropic. Elastic and plastic anisotropy of the
single crystal grains is supposed to result in an average
isotropic behavior within the considered material
volume containing a large number of grains. %Thus, linear isotropic
%elasticity \cite{Elasticity_Ciarlet_1} can be applied to describe
%the behavior of single-phased polycrystalline metal under external
%loading and plasticity can be taken into account by the method of
%Neuber shakedown \cite{Neuber}, for example.
The following probabilistic model for LCF is motivated by
simplicity and continuity in the sense that its basis is the
'deterministic' CMB life prediction approach. We furthermore
follow a simple scale-shape formulation for the probability law
that is quite common in reliability statistics, confer
\cite{Escobar_Meeker}, for example.

Let the scale field $N_{i_{\textrm{det}}}(\mathbf{x})$,
$\mathbf{x}\in\partial\Omega$, be the solution of the CMB equation
(\ref{materials_1.1}):
\begin{equation}\label{materials_5.1}
\varepsilon(\mathbf{x})=\frac{\sigma_f'}{E}(2N_{i_{\textrm{det}}}(\mathbf{x}))^{b}+\varepsilon_f'(2N_{i_{\textrm{det}}}(\mathbf{x}))^{c},
\end{equation}
where $\varepsilon(\mathbf{x})$ is the strain field. %computed
%from linear isotropic elasticity (which leads to a stress tensor
%$\sigma^e$) with subsequent application of appropriate
%stress-strain relationships also considering plasticity, confer
%\cite{Neuber}, \cite{Ramberg} and \cite{Harders_Roesler} for
%examples. If $\sigma$ is measured, plasticity is already included
%and FEA can be omitted to obtain $\varepsilon_a$.
Having obtained the scale field $N_{i_{\textrm{det}}}$
%calculated from $\varepsilon^e(x)$ via the stress-strain relation,
%calculation of the elastic von Mieses stress $\sigma_v^e$, the
%shake down and the Ramberg-Osgood formula.
we now follow a Weibull approach, see also \cite{Fedelich}, and
set for some shape parameter $m$
\begin{equation}
\label{eqaIntensityModel}
\rho(n,\varepsilon(x))=\frac{m}{N_{i_{\textrm{det}}}(\varepsilon(x))}\left(\frac{n}{N_{i_{\textrm{det}}}(\varepsilon(x))}\right)^{m-1}.
\end{equation}
Inserting this into (\ref{eqaDist}) and integrating over $n$, we
arrive at the cumulative distribution function for the proposed
probabilistic LCF model
\begin{equation}\label{materials_4.1}
F_{N_i}(n)=1-\exp\left(-\int_{\partial\Omega}\left(\frac{n}{N_{i_{\textrm{det}}}}\right)^{m}dA\right)
\end{equation}
for $n\geq0$ and some $m\geq1$, which yields the probability for
LCF crack initiation in the interval $(0,n]$. Note that
$N_{i_{\textrm{det}}}$
 has the units $[N_{i_{\textrm{det}}}]=$cycles$\times$ meter${}^{2/m}$ which is
achieved by changing the units of $\sigma_f'$ and $\epsilon_f'$
accordingly.

The shape parameter $m$ determines the scatter of the distribution
where small values for $m\geq1$ correspond\footnote{$0<m\leq1$ is
not realistic for fatigue.} to a large scatter and where the limit
$m\to\infty$ is the deterministic limit. Note that the approach
via (\ref{materials_4.1}) includes the assumption that $m$ is
independent of the strain state $\varepsilon$. The Weibull hazard
function can be easily replaced by any other differentiable hazard
function with scale parameter ${N_i}_{\textrm{det}}$. Furthermore,
it is important to stress that volume driven fatigue could be
considered as well by replacing the surface integral in
(\ref{materials_4.1}) with a volume integral whose integrand only
differs by different material parameters. For a discussion of
volume driven fatigue such as high-cycle fatigue HCF confer
\cite{Harders_Roesler,Vormwald}, for example.

Now, we show how to calibrate them by means of LCF test results
with standardized test specimens using of the maximum likelihood
method, confer \cite{Escobar_Meeker}. First, note that the
cumulative distribution function $F_{N_i}(n)$ of
(\ref{materials_4.1}) can be effectively simplified, as the
surface is subjected to homogeneous strain and temperature fields.
Thus, the surface integral reduces to multiplication with the
surface area between the gauge length. For the probability density
function $f_{N_i}(n)=\frac{d}{dn}F_{N_i}(n)$ with
$\eta=\left(\int_{\partial\Omega}N_{i_{\textrm{det}}}^{-m}dA\right)^{-1/m}$
the expression
\begin{equation}\label{results_1.0}
f_{N_i}(n)=\frac{m}{\eta}\left(\frac{n}{\eta}\right)^{m-1}\exp\left[-\left(\frac{n}{\eta}\right)^m\right]
\end{equation}
holds. We subsume the parameters (CMB parameters and the Weibull
shape parameter $m$) of the model in a vector $\theta$. Let
$\left\{n_i,\varepsilon_i,|\partial\Omega_i|\right\}_{i=1,\dots,q}$
denote the experimental data set for $q$ strain controlled LCF
tests, where $n_i$ is the number of cycles until crack initiation,
$\varepsilon_i$ the strain on the gauge surface and where
$|\partial\Omega_i|$ is the surface area in the gauge length. We
estimate $\theta$ by means of maximum likelihood, where the
so-called log-likelihood function
\begin{align}
\begin{split}
&\log\left(\mathcal{L}\left(\{(n_i,\varepsilon_i,|\partial\Omega_i|)\}_{i\in\{1,\dots,q\}}\right)[\theta]\right)=\sum_{i=1}^q\log(f_{N_i}(|\partial\Omega_i|,\varepsilon_i)(n_i)[\theta]).
\end{split}
\end{align}
is maximized with respect to the parameters. Thus, the likelihood
estimator $\hat{\theta}$ is given by
\begin{align}
\begin{split}
&\log\left(\mathcal{L}\left(\{(n_i,\varepsilon_i,|\partial\Omega_i|)\}_{i\in\{1,\dots,q\}}\right)[\hat{\theta}]\right)\\
=&\max_{\theta}\left\{\log\left(\mathcal{L}\left(\{(n_i,\varepsilon_i,|\partial\Omega_i|)\}_{i\in\{1,\dots,q\}}\right)[\theta]\right)\right\}.
\end{split}
\end{align}
%In this example case we use the optimization software package
%DAKOTA 5.0 to find the estimator $\hat{\theta}$.
Recalling that the CMB parameters are not the same as obtained
from fitting the standard CMB approach we will consider this fact
in more detail in the next subsection.

%%%%%%%%%%%%%%%%%%%%%%%%%%%%%%%%%%%%%%%%%%%%%%%%%%%%%%%%%%%%%%%%%%%%%%
\subsection{A New Interpretation of the CMB Parameters}\label{subsection_interpretation_CMB}

The CMB equation is a model for LCF life of standardized test
specimens. %The parameters of the CMB ansatz depend on the
%temperature field which is in LCF tests mostly chosen to be
%homogeneous. Thus, there are different W\"ohler curves for
%different temperatures.
However, due to the statistical nature of fatigue-crack initiation
the specimen size has an influence on crack-initiation life, i.e.
the number of cycles to crack initiation should decrease with
increasing specimen size. This effect is based on the assumption,
that the number of possible crack initiation sites increases with
specimen size, confer \cite{Kloos_Buch,Bazios_Gudladt}. As in our
case cracks are generally initiated at slip bands in surface
grains, the quantity that determines the size effect is the
specimen surface.
%as different surface areas of the specimen lead to different
%W\"ohler curves. Under the same conditions, a standardized test
%specimen with a greater surface area has a shorter LCF life
%because there are more grains on the surface subject to LCF failure mechanism.
In the following, we give a new interpretation of the CMB
parameters in context of the probabilistic model for LCF which
considers the size effect via the assumption of independently
occurring crack initiations. We also derive a relationship between
the parameters of the model and those of the original CMB
approach. This will also be important for the validation of the
model. %Moreover, we establish a relationship between the CMB
%parameters of specimens with different gauge surface areas by
%means of our probabilistic model for LCF.

At first, we employ the fact that the specimen is subject to
homogeneous temperature and strain fields at the gauge surface
$\partial\Omega$ which leads to
$\int_{\partial\Omega}\left(n/{N_{i_{\textrm{det}}}}\right)^{m}dA=|\partial\Omega|\left(n/
{N_{i_{\textrm{det}}}}\right)^{m}$ so that we can rewrite the
cumulative distribution function (\ref{materials_4.1}):
\begin{equation}\label{interpretation_CMB.1.1}
F_{N_i}(n)=1-\exp\left(-\left(\frac{n}{\eta_{\textrm{det}}(|\partial\Omega|)}\right)^{m}\right),
\end{equation}
where $|\partial\Omega|$ is defined as the area of the gauge
surface $\partial\Omega$ and where
\begin{equation}\label{interpretation_CMB.2.1}
\eta_{\textrm{det}}(|\partial\Omega|)=|\partial\Omega|^{-\frac{1}{m}}
N_{i_{\textrm{det}}}
\end{equation}
is the Weibull scale. Because a standardized test specimen with
gauge surface area equal to $1$ satisfies
$\eta_{\textrm{det}}(|\partial\Omega|=1)=N_{i_{\textrm{det}}}$
(from now on, all such specimens are called unit specimens),
$N_{i_{\textrm{det}}}$ is the Weibull scale of the unit specimen.
Note that $\eta_{\textrm{det}}(|\partial\Omega|)$ and
$N_{i_{\textrm{det}}}$ are the $1-\frac{1}{e}\approx63\%$
quantiles of crack initiation life as is the case for the scale
parameter of every Weibull distribution. Thus, we will also write
$\eta_{\textrm{det}}^{63\%}(|\partial\Omega|)$ and
$N_{i_{\textrm{det}}}^{63\%}$ instead of
$\eta_{\textrm{det}}(|\partial\Omega|)$ and
$N_{i_{\textrm{det}}}$, respectively. Note
that %the following relationship holds
\begin{equation}\label{interpretation_CMB.3.1}
%\eta_{\textrm{det}}(|\partial\Omega|)=
\eta_{\textrm{det}}^{63\%}(|\partial\Omega|)=(\ln2)^{-\frac{1}{m}}\,\,\eta_{det}^{50\%}(|\partial\Omega|),
\end{equation}
where $\eta^{50\%}_{\textrm{det}}$ is the $50\%$ quantile (median)
with respect to the Weibull distribution of the test specimen with
gauge surface area $|\partial\Omega|$. Our probabilistic model
assumes that the solution of the CMB equation yields
$N_{i_{\textrm{det}}}^{63\%}$. Therefore, one can interpret the
CMB parameters $b,c,\varepsilon',\sigma'$ to be belonging to the
W\"ohler curve of the unit specimen.

In many statistical methods of fatigue analysis the computed
W\"ohler curve yields a median value for the number of life
cycles. This value corresponds to
$\eta_{\textrm{det}}^{50\%}(|\partial\Omega|)$. Thus, we consider
$\eta_{\textrm{det}}^{50\%}(|\partial\Omega|)$ by combining
(\ref{interpretation_CMB.2.1}) and (\ref{interpretation_CMB.3.1})
\begin{equation*}
N_{i_{\textrm{det}}}=|\partial\Omega|^{\frac{1}{m}}\,\,\eta_{\textrm{det}}(|\partial\Omega|)=
\left(\frac{|\partial\Omega|}{\ln(2)}\right)^{\frac{1}{m}}\,\,\eta^{50\%}_{\textrm{det}}(|\partial\Omega|).
\end{equation*}
Inserting this expression into the previously mentioned CMB
equation for the unit specimen results in
\begin{align}\label{interpretation_CMB.4.1}
\varepsilon_a=\frac{\sigma_f'}{E}\left(\frac{|\partial\Omega|}{\ln(2)}\right)^{\frac{b}{m}}\left(2\,\eta^{50\%}_{\textrm{det}}(|\partial\Omega|)\right)^{b}+
\varepsilon_f'\left(\frac{|\partial\Omega|}{\ln(2)}\right)^{\frac{c}{m}}\left(2\,\eta^{50\%}_{\textrm{det}}(|\partial\Omega|)\right)^{c}.
\end{align}
Noting that the exponents $b$ and $c$ are not affected by the size
effect and defining
\begin{align}\label{interpretation_CMB.5.1}
\sigma_f'(|\partial\Omega|)=\left(\frac{|\partial\Omega|}{\ln(2)}\right)^{\frac{b}{m}}\sigma_f',\quad
\varepsilon_f'(|\partial\Omega|)=\left(\frac{|\partial\Omega|}{\ln(2)}\right)^{\frac{c}{m}}\varepsilon_f',
\end{align}
equation (\ref{interpretation_CMB.4.1}) leads to the following CMB
equation for the standardized specimen with gauge surface area
$|\partial\Omega|$:
\begin{align}\label{interpretation_CMB.6.1}
\varepsilon_a=\frac{\sigma_f'(|\partial\Omega|)}{E}\left(2\,\eta^{50\%}_{\textrm{det}}(|\partial\Omega|)\right)^{b}+
\varepsilon_f'(|\partial\Omega|)\left(2\,\eta^{50\%}_{\textrm{det}}(|\partial\Omega|)\right)^{c}.
\end{align}
Because the parameters $b,c,\varepsilon',\sigma'$ are a result of
fitting our probabilistic model the CMB equation
(\ref{interpretation_CMB.6.1}) is the prediction of our model for
the W\"ohler curve of a standardized specimen with gauge surface
area $|\partial\Omega|$. In the next section, we will validate
this prediction by fitting our model to an LCF test campaign and
comparing a predicted W\"ohler curve to the outcome of another
test campaign.

Vice versa, we can use already existing values for CMB parameters
of the original CMB approach to compute the CMB parameters of our
new model by means of (\ref{interpretation_CMB.5.1}): %Recall that
%we assume that the exponents $b$ and $c$ are not affected by the
%size effect.
%Finally, (\ref{interpretation_CMB.5.1}) results in a relationship
%between
For different gauge surface areas $|\partial\Omega_1|$ and
$|\partial\Omega_2|$ we obtain
\begin{align}\label{interpretation_CMB.7.1}
\frac{\sigma_f'(|\partial\Omega_1|)}{\sigma_f'(|\partial\Omega_2|)}=\left(\frac{|\partial\Omega_1|}{|\partial\Omega_2|}\right)^{\frac{b}{m}},\quad
\frac{\varepsilon_f'(|\partial\Omega_1|)}{\varepsilon_f'(|\partial\Omega_2|)}=\left(\frac{|\partial\Omega_1|}{|\partial\Omega_2|}\right)^{\frac{c}{m}}.
\end{align}
Equations (\ref{interpretation_CMB.5.1}) and
(\ref{interpretation_CMB.7.1}) show that for small $m\geq1$, i.e.
for significantly high scatter, the size effect plays an important
role, whereas in the deterministic limit $m\rightarrow\infty$
there is no size effect.

%%%%%%%%%%%%%%%%%%%%%%%%%%%%%%%%%%%%%%%%%%%%%%%%%%%%%%%%%%%%%%%%%%%%%%
\section{Experimental validation}\label{section_experiments}

In this section we consider LCF test results of specimens with
different geometries to calibrate and validate the proposed
probabilistic model. The specimens are made of a polycrystalline
superalloy. %We discuss fit and prediction results of our model for
%W\"ohler curves of the different specimen geometries.

%%%%%%%%%%%%%%%%%%%%%%%%%%%%%%%%%%%%%%%%%%%%%%%%%%%%%%%%%%%%%%%%%%%%%%
\subsection{Material and
Specimens}\label{subsection_material_specimen}

\begin{figure}[htbp]
  \centering
  \begin{minipage}[b]{0.4\textwidth}
    \includegraphics[width=5.5cm]{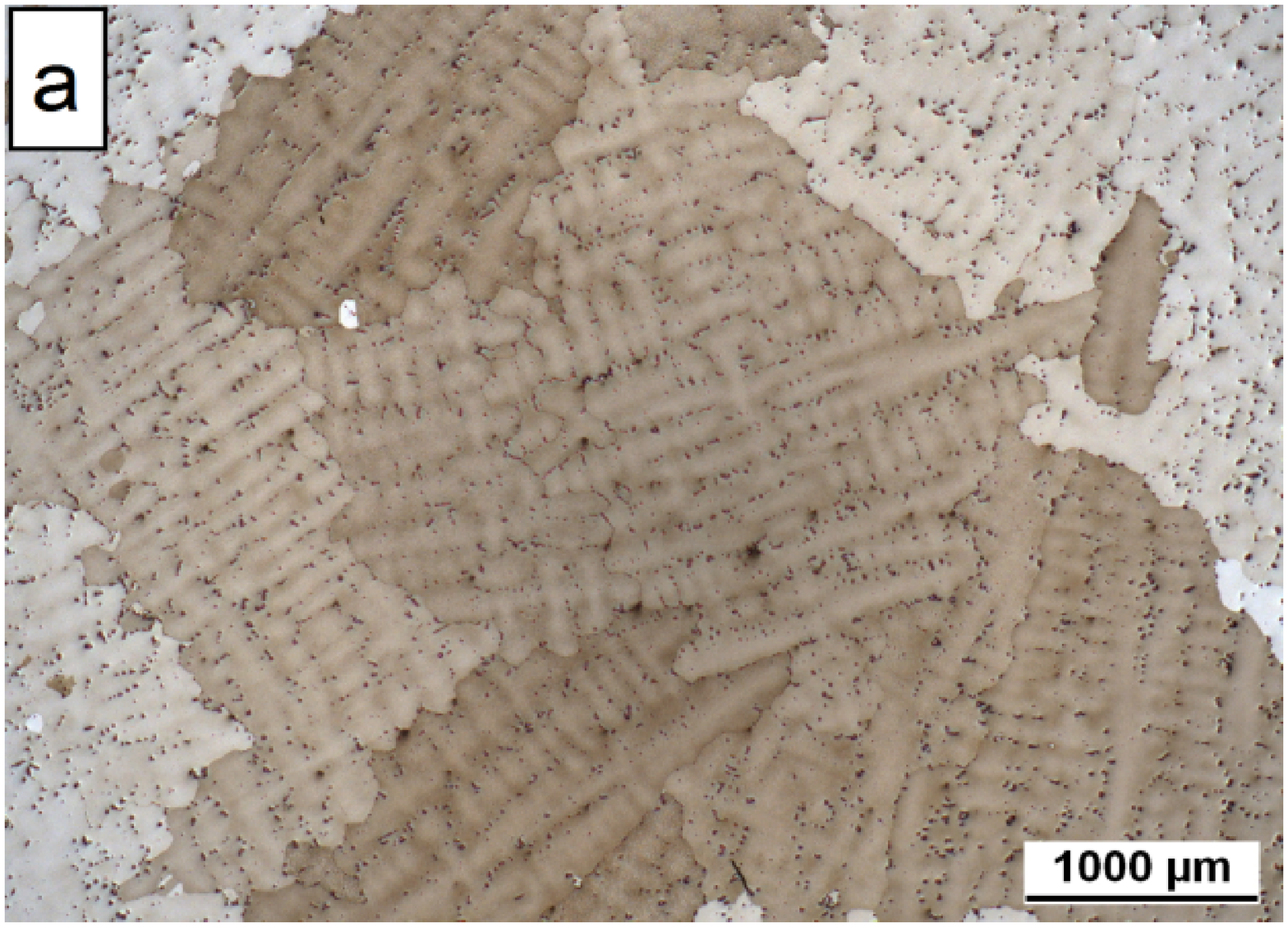}
  \end{minipage}
  \hfill
  \begin{minipage}[b]{0.4\textwidth}
    \includegraphics[width=5.5cm]{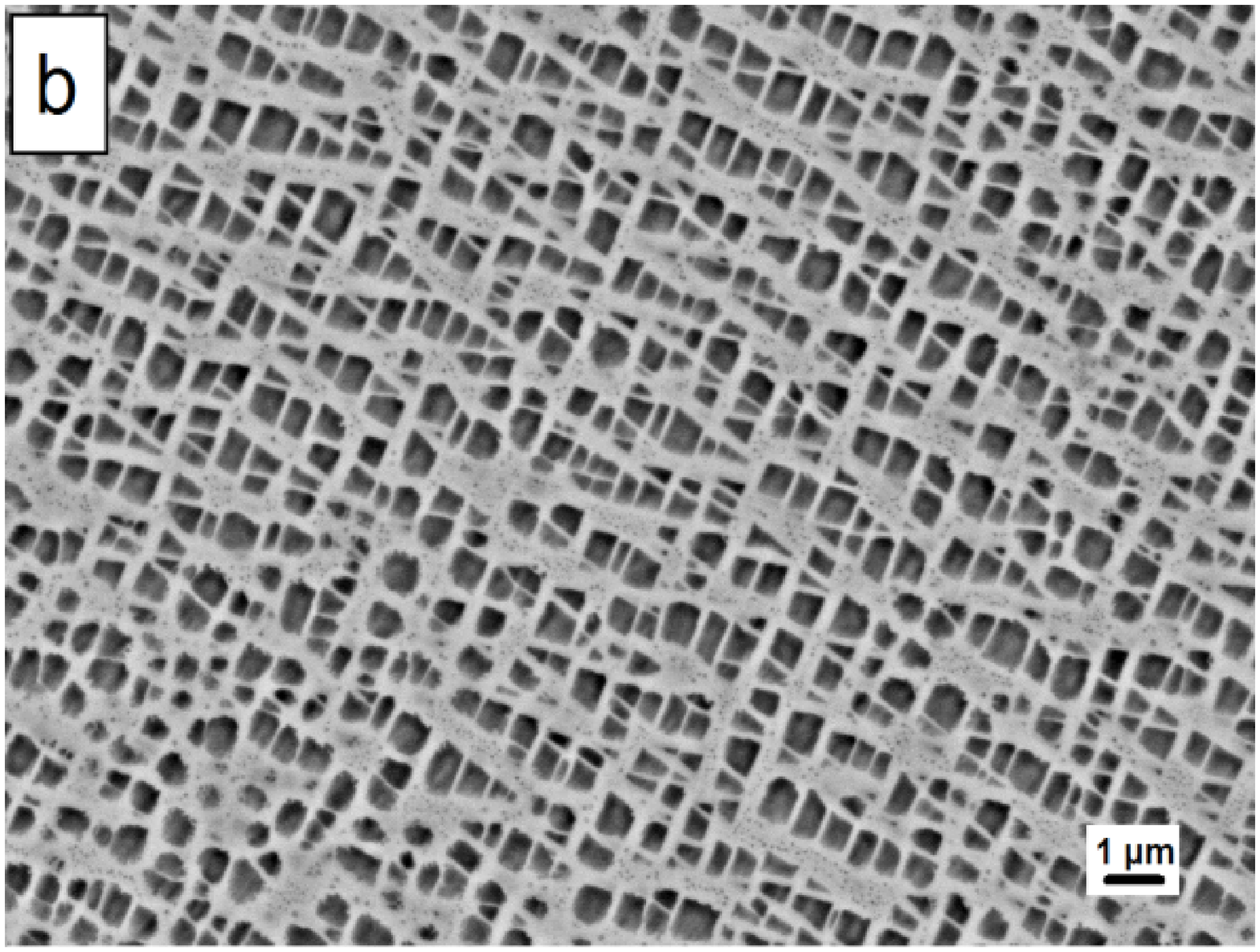}
  \end{minipage}
  \caption{\emph{Microstructure of RENE 80 by using OM (a) and REM (b).}}
  \label{Figure_rene80}
\end{figure}

The investigated material is the polycrystalline cast nickel base
superalloy RENE 80. The chemical composition of the alloy is given
in Table \ref{table_chemical_composition}.
%\vspace{5mm}
\begin{table}[t]
\centering% NICHT \begin{center}
\begin{tabular}{|c|c|c|c|c|c|c|c|c|c|c|}%\label{table.1.1}
  \hline
  Element & Ni & Cr & Co & Ti & Mo & W & Al & C & B & Zr \\
  \hline
  Weight - $\%$ & 60.0 & 14.0 & 9.5 & 5.0 & 4.0 & 4.0 & 3.0 & 0.17 & 0.15 & 0.03 \\
  \hline
\end{tabular}
\caption{Chemical composition of the investigated
material.}\label{table_chemical_composition}
\end{table}
%\vspace{5mm}
%The heat treated material was produced by Doncasters Group Ltd in
%the shape of cast slabs with dimensions 200 x 112 x 20mm.
In Figure \ref{Figure_rene80} the microstructure of the material
after heat treatment is shown in an optical (OM) and scanning
electron microscopy (SEM) image. The material shows the typical
dendritic structure of a cast alloy. Furthermore the
microstructure is quite coarse grained with a grain diameter of
approximately 2mm (Figure \ref{Figure_rene80}a) and strengthened
by ordered $\gamma'$ - precipitates, which appear in a cubic
morphology embedded in the $\gamma$ - matrix (Figure
\ref{Figure_rene80}b).

\begin{figure}[bhtp]
  \centering
  %\begin{minipage}[c]{7 cm}
    \includegraphics[width=11.5cm]{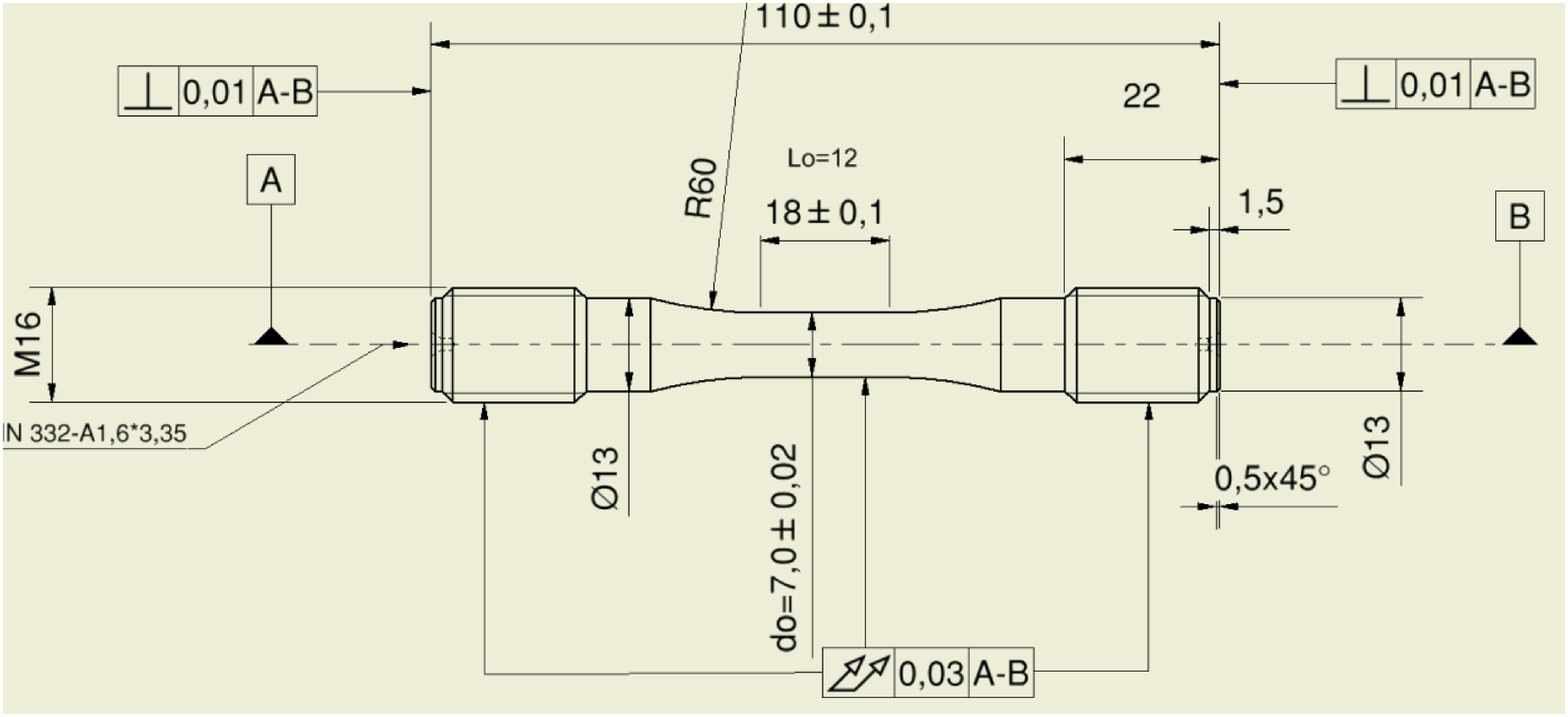}
  %\end{minipage}
  %\begin{minipage}[c]{7 cm}
  %  \includegraphics[width=8.5cm]{Specimen_large.eps}
  %\end{minipage}
  \caption{\emph{Small specimen geometry for the fatigue tests.}}
  \label{Figure_specimens}
\end{figure}

Cylindrical rods were eroded by electro-discharge-machining from
the slabs and afterwards turned to the final specimen geometry.
Figure \ref{Figure_specimens} shows one of the specimen geometries
used for the LCF tests. The diameter ($D_{\textrm{small}}$), the
gauge length ($L_{\textrm{small}}$) and the resulting surface and
volume of the gauge section of the specimen are listed in Table
\ref{table_specimen_geometry}.

%\vspace{5mm}

\begin{table}[t]
\centering% NICHT \begin{center}
\begin{tabular}{|c|c|c|c|}
  \hline
$L_{\textrm{small}}$ \textbf{[mm]} & $D_{\textrm{small}}$ \textbf{[mm]} & \textbf{Surface $\textrm{[mm}^2\textrm{]}$} & \textbf{Vol. $\textrm{[mm}^3\textrm{]}$} \\
  \hline
12.0 & 7.0 & 263.9 & 461.8 \\
  \hline
  %$G_2$ & 12 & 11 & 414.69 & 1140.39 \\
  %\hline
\end{tabular}
\caption{Geometrical parameters of the small specimen
geometry.}\label{table_specimen_geometry}
\end{table}

%\vspace{5mm}

In addition to tests with this specimen geometry, we consider
already existing LCF test results of Siemens AG for
T=850$\,^\circ$C regarding the same material subject to the same
heat treatment. The corresponding specimen geometry has a 2.86
times larger surface of the gauge section. In the following, we
refer to these specimens as the standard specimens and to the
specimens according to Table 2 as the small ones.

For the investigation of the LCF lifetime behavior a
servo-hydraulic fatigue testing machine with a maximum load of
100kN has been used. All LCF tests were carried out at
$850\,^\circ$C at isothermal conditions in total strain control
with a load ratio\footnote{The load ratio is defined by the
minimum strain divided by the maximum strain.} of $R=-1$. The load
cycles were applied with triangular waveform at a frequency of
0.1Hz for high strain amplitudes ($\varepsilon_{a,\textrm{tot}} =
0.55\% - 0.65\%$) and $1$Hz for low strain amplitudes
($\varepsilon_{a,\textrm{tot}} = 0.15\% -
0.25\%$), respectively. %The results of both test campaigns is shown in Figure
%\ref{Figure_woehler_small_standard} and will be discussed in the
%next subsection.

The crack-initiation lifetime is defined by a drop of the curve
which is given by the maximum stress' dependence on the number of
already conducted load cycles. In case of no extremely large
strain amplitude, this curve starts with approximately stable
maximum stress before first crack initiations take place after a
certain number of load cycles. When the first crack initiation
occurs, the curve starts to drop as the cross-section area of the
specimen decreases and thereby a lower tensile force is needed for
imposing the prescribed strain amplitude. Note that the maximum
stress is defined as the ratio of this tensile force (which
depends on the current load cycle) and the fixed cross-section
area of the specimen at the beginning of the LCF tests.
%at the beginning compared to the stabilized of the cyclic deformation curves.
%The crack-initiation lifetime was defined
%by a drop of the maximum stress compared to the stabilized of the
%cyclic deformation curves.
Considering equal crack areas at crack initiation life $N_i$, the
drop of the maximum stress was calculated separately for each
geometry using the following equation:
\begin{equation}
S_{\textrm{small}}
=S_{\textrm{standard}}\cdot\left(\frac{D_{\textrm{standard}}}{D_{\textrm{small}}}\right)^2
\end{equation}
with $S_{\textrm{small/standard}}$ stress drops and
$D_{\textrm{small/standard}}$ diameters of the small and standard
specimen, respectively. %In this work, the number of cycles to
%failure was defined for a maximum stress drop of $5\%$ for the
%standard and of $10\%$ for the small specimen which corresponds to
%a crack size of $3.926\textrm{mm}^2$.
The LCF test results for both specimen geometries are given in
Figure \ref{Figure_woehler_small_standard} in the representation
total strain amplitude ($\varepsilon_{a,\textrm{tot}}$) versus
cycles to crack initiation ($N_i$).

%%%%%%%%%%%%%%%%%%%%%%%%%%%%%%%%%%%%%%%%%%%%%%%%%%%%%%%%%%%%%%%%%%%%%%
\subsection{Calibration of the Model and Prediction of W\"ohler Curves}\label{subsection_calibration}

\begin{figure}[htbp]
  \centering
    \includegraphics[width=11.75cm]{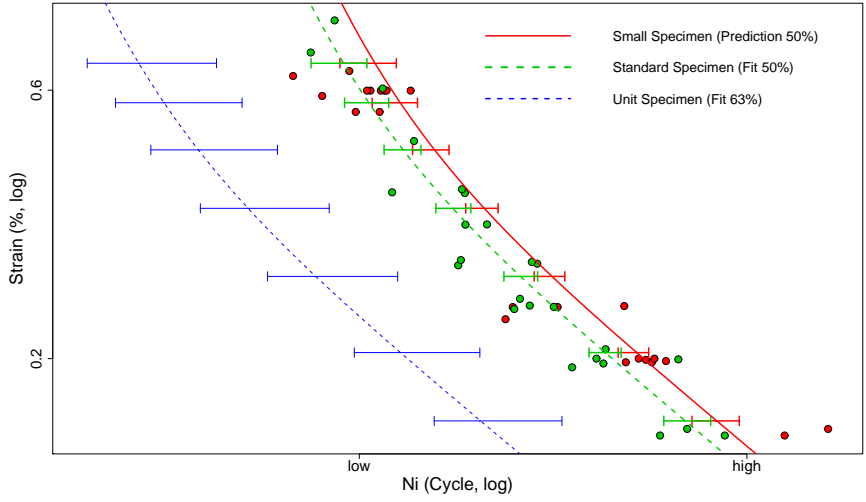}
  \caption{\emph{LCF test results of the small (red) and standard (green) specimens.
  The standard specimen data are used for the calibration
  of the model. The calibration leads to the W\"ohler curve (blue) of the
  unit specimen and, via formulae (\ref{interpretation_CMB.5.1}), to the
  W\"ohler curve (green) of the standard specimen. The W\"ohler curve (red)
  for the small specimen is predicted by the probabilistic model for LCF,
  where formulae (\ref{interpretation_CMB.5.1}) is employed, again}.}
  \label{Figure_woehler_small_standard}
\end{figure}

For the calibration of the model we have chosen the LCF test
results $N_{i_\textrm{standard}}(\varepsilon_{a,\textrm{tot}})$ of
the standard specimen geometry as the corresponding data basis is
much larger here than for the small specimen. In Figure
\ref{Figure_woehler_small_standard}, test results for both
geometries are shown, together with several W\"ohler curves.
%In this subsection we show the results of the calibration of the
%probabilistic model with respect to the LCF test results
%$\textrm{Ni}_\textrm{standard}(\varepsilon_{a,\textrm{tot}})$ of
%the standard specimen geometry. Then, we use our probabilistic
%model for LCF to predict the W\"ohler curve for the small specimen
%geometry. Figure \ref{Figure_woehler_small_standard} shows the LCF
%test results for both specimen geometries. Since the number of
%significantly different strain amplitudes is much higher for the
%standard specimen than for the small one we employ its test data
%for the calibration of our probabilistic model.
The calibration of the probabilistic model for LCF is conducted
according to the maximum likelihood method as described at the end
of Subsection \ref{subsection_stochastics}. The W\"ohler curve of
the unit (blue) specimen describes the functional dependence of
the Weibull scale parameter on the strain amplitude
$\varepsilon_{a,\textrm{tot}}$. Using formulae
(\ref{interpretation_CMB.5.1}) leads to the W\"ohler curves of the
standard (green) and small (red) specimens with the 50$\%$-Weibull
quantiles (medians). It is important to point out that the curves
for the unit and standard specimens are based on the calibration,
whereas the curve for the small specimen is a pure prediction of
our probabilistic model. The horizontal bars denote the 92.5$\%$
confidence intervals -- see Figure
\ref{Figure_woehler_small_standard} -- that have been computed via
a fully parametric bootstrap sampling procedure in conjunction
with the percentile method as described in
\cite{Escobar_Meeker,Davison_Hinkley}.
%The level of significance is 7.5$\%$.

%\begin{figure}[htbp]
%  \centering
%  \begin{minipage}[c]{7 cm}
%    \includegraphics[width=7cm]{Q_diagnostic_1.eps}
%  \end{minipage}
%  \begin{minipage}[c]{7 cm}
%      \includegraphics[width=7cm]{Q_diagnostic_1.eps}
%%    \scalebox{0.27}{\input{Sampling_large_specimen_1.pstex_t}}
%    %\includegraphics[width=8.5cm]{Sampling_large_specimen_1.pstex_t}
%  \end{minipage}
%  \caption{\emph{Quotient Q at the tested strain amplitudes (a)
%  and the large specimen (b).}}
%  \label{Figure_diagnostic_large_specimen}
%\end{figure}

\begin{figure}[htbp]
  \centering
      \includegraphics[width=14cm]{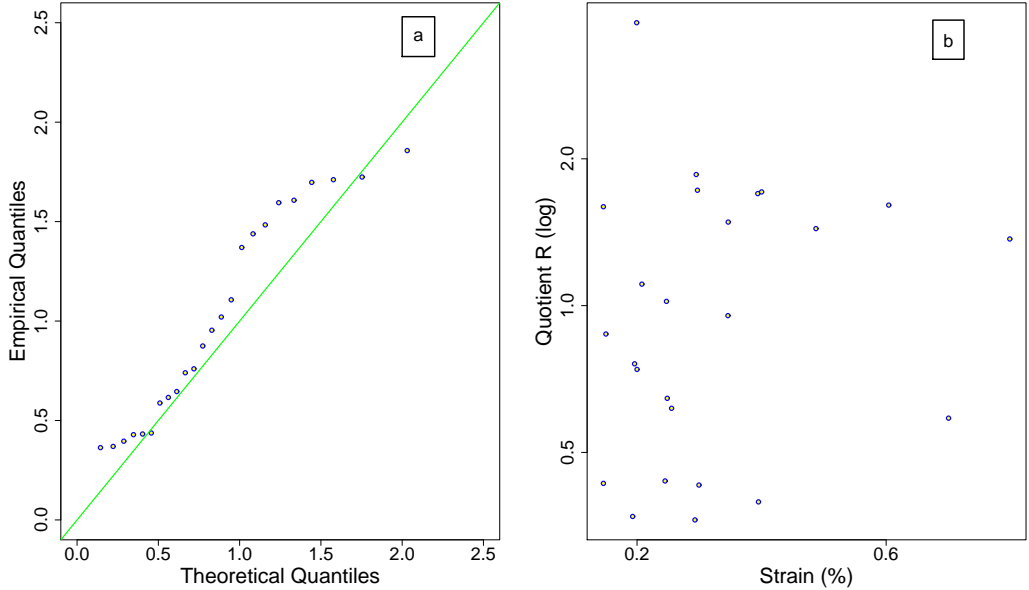}
  \caption{\emph{Q-Q plot (a) with respect to a Weibull fit of quotient $q$ with scale $1$ and
  values of $q$ (b) at the tested strain amplitudes of the standard specimen.}}
  \label{Figure_diagnostic_large_specimen}
\end{figure}

%model diagnostics:  semi-parametric multivariate investigations
%more data nachgeordneter effect

For model diagnostics we consider the quotient
\begin{equation}\label{exp_quotient}
q=\frac{N_{i_\textrm{standard}}(\varepsilon_{a,\textrm{tot}})}{N_{i_\textrm{det,standard}}(\varepsilon_{a,\textrm{tot}})},
\end{equation}
with $N_{i_\textrm{standard}}(\varepsilon_{a,\textrm{tot}})$ the
actual test results of the standard specimens and
$N_{i_\textrm{det,standard}}(\varepsilon_{a,\textrm{tot}})$ the
corresponding fit of our model for the Weibull scale parameter.
Figure \ref{Figure_diagnostic_large_specimen} (a) shows a Q-Q plot
with respect to a Weibull fit of $q$ with scale $1$. According to
the Kolmogorov - Smirnov test with a p-value of 17$\%$, deviations
from the Weibull distribution are not statistically significant
which supports the Weibull approach (\ref{eqaIntensityModel}) of
our model. Recalling our assumption that the shape parameter $m$
does not depend on the strain state $\varepsilon$, we consider
Figure \ref{Figure_diagnostic_large_specimen} (b) which shows the
values of $q$ at the tested strain levels of the standard
specimen. Statistical tests cannot exclude a deterministic
relationship between $q$ and $\varepsilon_{a,\textrm{tot}}$, in
particular for higher strain ranges, which could be finally
assessed with additional test results. Nevertheless, note that the
assumption of no significantly deterministic relationship between
those quantities is often used, in particular with respect to
design relevant strain ranges.

Because the unit specimen has the largest gauge surface of all
considered specimens its W\"ohler curve is expected to contain the
lowest life-cycle numbers $N_i$. This expectation is confirmed by
the results shown in Figure \ref{Figure_woehler_small_standard}.
The figure also shows that the fitting procedure and formulae
(\ref{interpretation_CMB.5.1}) are able to yield an appropriate
W\"ohler curve for the standard specimen. The predicted W\"ohler
curve for the small specimen is shifted to higher life-cycle
numbers $N_i$ due to the incorporated size effect in our
model\footnote{The model's main assumption of independently
occurring LCF crack initiations directly results in consideration
of the size effect.}. However, this predicted shift is not very
large so that statements on the position of the curves compared to
the test results have to be considered carefully. Nevertheless,
the LCF test results support the assumption of a size effect for
lower strain amplitudes. A size effect is more difficult to be
recognized in the range of higher strain amplitudes, where
plasticity effects are present. Comparing the position of the
W\"ohler curve and the LCF test results of the small specimen
shows that our model estimates the size effect for lower strain
amplitudes very well.
In the range of higher strain amplitudes, where plasticity comes
into play, our model seems to overestimate the shift to higher
life-cycle numbers $N_i$. Note that the prediction of our model in
that range has to be judged more carefully as only few calibration
data were available in the plastic range. This is also affirmed by
the confidence intervals which overlap more significantly in the
plastic range. Moreover, consider the large scatter in the data
and a variety of possible error sources in the experiments such as
determining crack initiation, stress and strain amplitudes and
slight deviations from the homogeneity of strain and temperature
fields.

Further investigations on the size effect can be found in
\cite{Vormwald,Bennett_McDowell,Dunne_Manonukul}, where a smaller
size effect for higher strain amplitudes is stated as well.
According to \cite{Bennett_McDowell,Dunne_Manonukul} the small
difference in the size effect between high and low strain
amplitudes may be explained by the fact that at lower strain
amplitudes plastic deformation is concentrated in individual slip
systems of favorably orientated grains, whereas at higher strain
amplitudes plastic deformation takes place over the whole gauge
length. According to \cite{Duyi_Dehai} this results in an
increased number of activated slip bands, so that a possible
effect of surface grain orientation becomes less important.

%%%%%%%%%%%%%%%%%%%%%%%%%%%%%%%%%%%%%%%%%%%%%%%%%%%%%%%%%%%%%%%%%%%%%%
\section{Conclusions}\label{section_conclusion}

A probabilistic model for LCF has been derived from concepts of
fatigue analysis and point processes. It was shown that this model
leads to a new interpretation of the CMB parameters which are
independent of the geometry of standardized test specimens.
Moreover, we have derived formulae for the prediction of W\"ohler
curves of standardized specimens with different surface areas. The
prediction of such W\"ohler curves was also the basis for the
validation of our probabilistic approach. % after we had calibrated
%the model with LCF test results.

The calibration of our model resulted in an appropriate fit for
the W\"ohler curve of the standard specimen. For strain amplitudes
not too high, we could find a size effect regarding the test
results of the small specimen, and our model was able to
appropriately predict the size effect's impact on the W\"ohler
curve. But for higher strain amplitudes in the plastic range,
the size effect could not be confirmed without ambiguity. %which is
%supported by results described in \cite{Vormwald}, for example.
In this range our model seems to overestimate a size effect.
However, considering the large scatter for LCF life and a variety
of possible error sources regarding the LCF tests, this
overestimation could not be finally concluded.

%In order to consider specimen geometries with a much larger
%difference in their gauge surface we conducted a validation using
%a sampling approach based on existing LCF data of the standard
%specimen. This approach enabled us to investigate whether our
%model is capable of predicting the impact of the size effect on
%the W\"ohler curves. It turned out that the model is indeed able
%to predict the W\"ohler curve of the small specimen accurately on
%the whole tested strain range.

For future work, we propose to analyze the size effect in the
plastic range in more detail. Recall that for larger values of
$m$, i.e. for smaller scatter, the size effect decreases. More LCF
test results in the plastic range could lead to a different value
of $m$. Furthermore, one could change the Weibull hazard approach
according to another distribution such as log-normal distribution.
%By means of censoring the maximum likelihood fit of our model
%according to design relevant LCF data one could hope for a larger
%value of $m$.
Furthermore, it might be worthwhile to incorporate a mathematical
concept for the slip systems of the grains which could lead to a
smaller scatter of the corresponding LCF results as the random
behavior of the slip systems is then taken into account.

The comparison of predicted W\"ohler curves for notched specimens
with corresponding LCF test results will play an important role in
future work because this will show how effective the proposed
model will be in case of inhomogeneous strain fields. It then
might turn out that the model needs to incorporate local
information on the strain gradient. Similarly, the aspect of
inhomogeneous temperature fields can be taken into account in the
model if the CMB approach considers an appropriate deterministic
temperature model for LCF. The model can also consider volume
integrals and can thereby be applied to volume driven LCF failure
mechanism with initial creep damage and to HCF failure mechanism,
for example. %Moreover, applications and possible extensions of our
%model to thermal-mechanical fatigue (TMF) and non-stationary FEA
%could be worthwhile investigating.

The model is also intended to be applied to FEA simulations of
engineering parts under cyclic loading.
%The prediction for first LCF crack initiations
%could then be compared to corresponding field data such that a
%further validation of our model could be realized.
These simulations in conjunction with field data can be employed
to validate or recalibrate the proposed model. Finally, let us
mention an improved link to shape optimization which is given by
our probabilistic model. Due to the probabilistic nature of our
model the new objective functional (\ref{materials_4.1}) for LCF
life is sufficiently regular so that efficient gradient based
shape optimization schemes can be conducted,
confer \cite{Sokolowski_Zolesio,Shape_Optimization_Haslinger}. %In contrast to this
%ansatz the deterministic safe-life approach for fatigue design has
%a much less regular objective functional as the position of
%maximal strain is considered.

%%%%%%%%%%%%%%%%%%%%%%%%%%%%%%%%%%%%%%%%%%%%%%%%%%%%%%%%%%%%%%%%%%%%%%
\section*{Acknowledgments} This work has been supported by the
German federal ministry of economic affairs BMWi via an AG Turbo
grant. We thank the gas turbine technology department of Siemens
AG for stimulating and helpful discussions.

%%%%%%%%%%%%%%%%%%%%%%%%%%%%%%%%%%%%%%%%%%%%%%%%%%%%%%%%%%%%%%%%%%%%%%

%%%%%%%%%%%%%%%%%%%%%%%%%%%%%%%%%%%%%%%%%%%%%%%%%%%%%%%%%%%%%%%%%%%%%%



%% References


%\bibliographystyle{elsarticle-num}



\end{document}